\documentclass[a4paper,fleqn,usenatbib]{mnras}

\usepackage{newtxtext,newtxmath}

\usepackage[T1]{fontenc}
\usepackage{ae,aecompl}

\usepackage{graphicx}	
\usepackage{amsmath}	
\usepackage{amssymb}	

\newcommand{\apostle}{{\small\rm APOSTLE}}
\newcommand{\eagle}{{\small\rm EAGLE}}
\newcommand{\dove}{{\small\rm DOVE}}
\newcommand {\hi} {{\rm H}\,{\small\rm I}}

\newcommand {\hicap} {{\rm H}\,{\scriptsize\rm I}}

\newcommand {\kms} {\,{\rm km\,s}^{-1}}

\newcommand {\pc} {\,{\rm pc}}

\newcommand {\kpc} {\,{\rm kpc}}
\newcommand {\Mpc} {\,{\rm Mpc}}

\newcommand {\kmskpc} {\,{\rm km\,s}^{-1}\,{\rm \kpc}^{-1}}

\newcommand {\de}{^{\circ}}

\newcommand {\msun}{\,{\rm M}_\odot}

\newcommand{\Gyr}{\,{\rm Gyr}}

\newcommand {\msunyr}{\,{{\rm M}_\odot\,\rm yr}^{-1}}

\newcommand {\vlos}{V_{\rm LOS}}

\def\aj{AJ}%
\def\araa{ARA\&A}%
\def\apj{ApJ}%
\def\apjl{ApJ}%
\def\apjs{ApJS}%
\def\ao{Appl.~Opt.}%
\def\aap{A\&A}%
\def\aapr{A\&A~Rev.}%
\def\mnras{MNRAS}%
\title[Bars in dark matter-dominated dwarf galaxy discs]{Bars in dark matter-dominated
  dwarf galaxy discs}

\author[A. Marasco et al.]{
A. Marasco,$^{1}$\thanks{marasco@astron.nl} 
K. A. Oman,$^{2}$ 
J. F. Navarro,$^{2,3}$ 
C. S. Frenk$^{4}$
and T. Oosterloo$^{1,5}$
\\
$^{1}$ ASTRON, Netherlands Institute for Radio Astronomy, Postbus 2, 7900 AA, Dwingeloo, The Netherlands\\
$^{2}$ Department of Physics \& Astronomy, University of Victoria, Victoria, BC, V8P 5C2, Canada\\
$^{3}$ Senior CIfAR Fellow\\
$^{4}$ Institute for Computational Cosmology, Department of Physics, University of Durham, South Road, Durham DH1 3LE, United Kingdom\\
$^{5}$ Kapteyn Astronomical Institute, University of Groningen, Postbus 800, 9700 AV Groningen, The Netherlands
}

\date{Accepted XXX. Received YYY; in original form ZZZ}

\pubyear{2017}

\begin{document}
\label{firstpage}
\pagerange{\pageref{firstpage}--\pageref{lastpage}}
\maketitle

\begin{abstract}
  We study the shape and kinematics of simulated dwarf galaxy discs in
  the \apostle\ suite of $\Lambda$CDM cosmological hydrodynamical
  simulations. We find that a large fraction of these gas-rich,
  star-forming discs show weak bars in their stellar component,
  despite being dark matter-dominated systems. The bar pattern shape
  and orientation reflect the ellipticity of the dark matter
  potential, and its rotation is locked to the slow figure rotation of
  the triaxial dark halo.  The bar-like nature of the potential
  induces non-circular motions in the gas component, including
  strong bisymmetric flows that can be readily seen as $m\!=\!3$
  harmonic perturbations in the \hi\ line-of-sight velocity
  fields. Similar bisymmetric flows are seen in many galaxies of the
  THINGS and LITTLE THINGS surveys, although on average their
  amplitudes are a factor of $\sim 2$ weaker than in our simulated
  discs. Our results indicate that bar-like patterns may arise even
  when baryons are not dominant, and that they are common enough to
  warrant careful consideration when analyzing the gas kinematics of
  dwarf galaxy discs.
\end{abstract}




\section{Introduction}
\label{SecIntro}

Bars are a common morphological feature of spiral galaxies in the
local Universe.  About one-quarter of disc galaxies show strong bars
\citep{Masters+11,Cheung+13}, and this fraction increases up to about
two-thirds when including weaker features
\citep{MulchaeyRegan97,KormendyKennicutt04}.  The fraction of barred
galaxy appears to depend significantly on galaxy luminosity, being
lower for fainter galaxies \citep{MendezAbreu+10,Janz+12}, and, to a
minor extent, on environment, increasing in regions of higher galaxy
density \citep{MendezAbreu+12}.

Bars have been a topic of interest for decades, and numerical
simulations have played a major role in our current understanding of
their formation and evolution. Although details need to be fully
worked out, there is widespread agreement that bars develop in stellar
discs as a consequence of the inevitable tendency of collisionless
systems to evolve dynamically by redistributing mass inwards and
angular momentum outwards.

Bar patterns offer a particularly efficient way of carrying out this
redistribution, and grow and strengthen rapidly in cold, massive
stellar discs \citep[see, e.g.,][for recent
reviews]{Athanassoula13,Sellwood14}. Their growth is so rapid in such
systems \citep{MillerPrendergast68,Hockney+69} that few can avoid becoming `bar
unstable' in a couple of rotation periods \citep{SellwoodWilkinson93},
unless stabilized by the presence of a dominant dark halo
\citep{OstrikerPeebles73}.  These results have been so influential that
the presence of bar-like patterns is often taken to indicate discs
where the dark halo is gravitationally unimportant; i.e., discs that
are `maximal'.

More recent work, however, suggests a more complex scenario where bar
formation is delayed, but not fully inhibited, in discs where dark
matter haloes are gravitationally dominant
\citep[e.g.,][]{Athanassoula02,Athanassoula03,Algorry+17}: bars are
thus not synonymous with maximal discs, suggesting that care must be
exercised when using morphological features to infer indirectly the relative
importance of disc and halo in spiral galaxies.

Bars are not the only bisymmetric ($m=2$) perturbation expected in
disc galaxies. Indeed, discs are expected to form at the centre of
cold dark matter (CDM) haloes, which have long been known to be triaxial in
nature \citep{Frenk+88,Warren+92,JingSuto02}. Discs that settle on the
symmetry plane of such haloes would be subject to gravitational forces
akin to those in barred systems, although the bisymmetric pattern in
this case is expected to rotate much less rapidly than in typical
strongly-barred systems.

The consequences of halo triaxiality on the kinematics of discs have
not been extensively studied \citep[although see][for an
attempt]{Hayashi+07}, in part because the assembly of the disc is
thought to `sphericalise' the halo \citep[see][and references
therein]{Abadi+10}. However, this is only true in the case of massive discs
such as those of luminous, high surface brightness spirals. In the
case of fainter, lower surface brightness spirals, discs are less
gravitationally important and the sphericalisation of their
surrounding haloes should be less complete \citep{MachadoAthanassoula10,Kazantzidis+10}.


Regardless of origin, bar-like perturbations can have important
consequences on the kinematic of discs, particularly on the
interpretation of azimuthally-averaged rotation curves and of 2D
velocity fields of gas and stars. \citet[][hereafter HN06]{HayashiNavarro06}, for
example, showed that even minor deviations from spherical symmetry can
induce large deviations from circular orbits in the velocity field of
a gaseous disc.  This possibility was studied by
\citet{Trachternach+08}, who carried out a harmonic decomposition of
the \hi\ velocity fields in 19 galaxies from The \hi\ Nearby Galaxy
Survey \citep[THINGS,][]{Walter+08}. The harmonic decomposition
suggests that the magnitude of non-circular motions in the central
regions of these systems is small, and compatible with a
circularly-symmetric potential.

However, most of the systems studied by Trachternach et al. were large
spirals, and the analysis has not been extended to the dwarf galaxy
regime, where the triaxiality of the halo might be better preserved.
A harmonic decomposition of LITTLE THINGS galaxies \citep{Hunter+12} was performed by \citet{Oh+15}, but
their findings were not discussed in terms of triaxiality of the gravitational potential, thus whether or not dwarf galaxies 
are consistent with hosting triaxial halos is still an open question.

Recently, \citet[][hereafter O17]{Oman+17} have studied the gas
kinematics in a sample of 33 simulated dwarf galaxies from the
\apostle\ suite of $\Lambda$CDM cosmological simulations
\citep{Sawala+15,Fattahi+16}.  Their approach used synthetic \hi\
observations of these systems, from which the gas rotation curve was
derived using the same tilted ring modelling technique adopted in most
\hi\ observational studies.  

One of the main results of O17 is that, depending on the orientation of the
line of sight, a large variety of rotation curves might be derived for
the same system, even for fixed inclination.  The variety is due to
non-circular motions in the gas; in particular, to strong bisymmetric
($m=2$ harmonic) fluctuations in the azimuthal \hi\ velocity field,
which the tilted-ring model is not well suited to account for.  We
study here the origin of these non-circular motions, focussing on the
mass distribution and the shape of the gravitational potential. 
We show that the cause of these bisymmetric gas flows is the presence of bar-like features in both the stellar and the dark matter distributions.

This paper is structured as follows.  In Section \ref{apostle_summary}
we summarise the main features of the \apostle\ simulations and of the
selected galaxy sample.  In Section \ref{Results} we discuss the
properties of the stellar and dark matter bars in our sample, and how
they affect the gas kinematics.  A comparison with dark-matter-only
simulations and with the observations is also presented.  We finally summarise our findings in Section
\ref{Conclusions}.

\section{Simulations and sample selection}\label{apostle_summary}
A detailed description of the \apostle\footnote{A Project Of Simulating The Local Environment} simulations can be found in \citet{Sawala+15} and \citet{Fattahi+16}.
Here, we summarise their main characteristics.

\apostle\ is a suite of cosmological hydrodynamical simulations
performed in a $\Lambda$CDM framework, adopting the cosmological
parameters inferred from WMAP-7 data \citep{Komatsu+11}.  It consists of
12 subvolumes selected from the \dove\ cosmological simulation
\citep{Jenkins+13} and re-simulated using the zoom-in technique
\citep{Frenk+96,Power+03}.  The volumes are centred around two massive
haloes (analogous to the Milky Way - M31 pair) and chosen to resemble
these galaxies in terms of mass, separation and kinematics, whilst ensuring
relative isolation from more massive structures.

The hydrodynamics and the baryonic subgrid physics implemented in
\apostle\ are the same as those adopted in the \eagle\ simulations
\citep{Schaye+15,Crain+15}.  \eagle, and by extension \apostle, uses a
formulation of the smoothed-particle hydrodynamics (SPH) known as
{\small ANARCHY} \citep[Dalla Vecchia in prep, see also Appendix A
of][]{Schaye+15}, which alleviates significantly the issues related to
artificial gas clumping and the poor treatment of hydrodynamical
instabilities associated with the classical SPH scheme
\citep[e.g][]{Kaufmann+06,Agertz+07}.  Other features introduced by
{\small ANARCHY} are the use of an artificial viscosity switch
\citep{CullenDehnen10}, an artificial conduction switch
\citep{Price08} and a timestep limiter \citep{DurierDallaVecchia12}.
The recipes for subgrid physics include the star formation
implementation of \citet{SchayeDallaVecchia08}, thermal star formation
feedback from \citet{DallaVecchiaSchaye12}, radiative gas cooling and
photo-heating from \citet{Wiersma+09a} and stellar mass loss from
\citet{Wiersma+09b}.  Accretion onto black holes and AGN feedback is not implemented in
\apostle\ but this is not critical for the mass scale of interest.

The \apostle\ volumes are resimulated at three levels of resolution.
As in O17, here we focus on the highest resolution level \citep[`L1'
in][]{Fattahi+16}, featuring a dark matter particle mass of
$3.6\times10^4\msun$, a gas particle initial mass of
$7.4\times10^3\msun$ and maximum softening length of $134\pc$.  At the
moment of writing, only five of the 12 volumes have been re-simulated
at this resolution.  Galaxies, and in general `subhaloes', are
identified in the simulation via the {\small SUBFIND} algorithm
\citep{Dolag+09}, which is based on a friend-of-friends (FoF) scheme.

O17 studied 33 dwarf systems, selecting central\footnote{A
  `central' subhalo is the most massive subhalo of each FoF group, and
  is therefore not a satellite of a more massive system.} subhaloes
with maximum circular velocity $60\!<\!V_{\rm max}\!<\!120\kms$ at
redshift $z\!=\!0$.  The $V_{\rm max}$ selection targets the `dwarf'
spiral regime, while the exclusion of satellites allows us to discard
systems whose features might be affected by the effect of
tides.  

We focus here on the same galaxy sample, whose main properties are
listed in Table A1 of O17.  Each simulated galaxy is uniquely
identified by four labels: the resolution level (always L1 in our sample), the \apostle\ volume it belongs to (V1,
V2,..., V12), the FoF group, and {\small SUBFIND} sub-group
numbers, respectively.  The latter is always equal to 0, indicating
that the system is the central subhalo of its FoF group.
These labels follow the `AP' keyword that identifies the suite of simulations used.

As in other studies of neutral hydrogen using the \eagle\ simulations,
the \hi\ mass fraction of each gas particle in the \apostle\ volumes
is computed in two steps: the fraction of neutral hydrogen is first
derived following the prescription of \citet{Rahmati+13} for
self-shielding from the metagalactic ionising background, and then a
pressure-dependent correction for the molecular gas fraction is
applied following \citet{BlitzRosolowsky06}. We refer the interested
reader to \citet{Crain+17} for details.

\section{Results}\label{Results}
\begin{figure*}
\includegraphics[width=\textwidth]{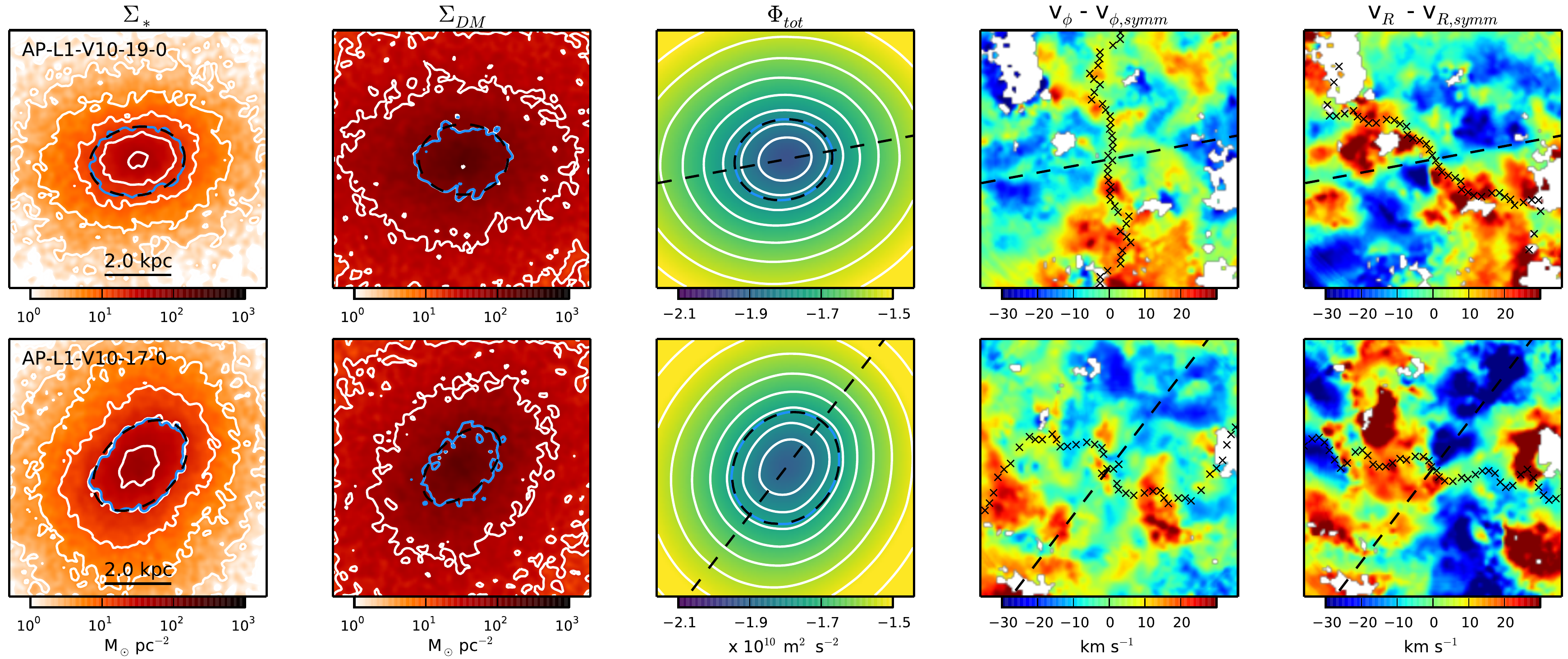}
\caption{Face-on maps of two representative \apostle\ dwarfs at redshift $0$. \emph{First and second columns:} stellar and dark matter surface density maps, shown on the same colour scale. Isodensity contours (in white) are spaced by 0.3 dex. The blue isodensity contour has semi-major axis equal to the projected radius at half stellar mass, $R_{\rm h}$; the dashed black line shows the best-fit ellipse to this contour. \emph{Third column:} gravitational potential map on the galactic midplane. \emph{Fourth and fifth columns:} maps of the azimuthal and radial \hicap\ velocity fields, also on the midplane, from which we have subtracted the mean velocity computed at different radii. A residual $m\!=\!2$ mode is clearly visible in both maps; its phase is traced by black crosses. The dashed black line in the rightmost three panels shows the major axis of the gravitational potential.}
\label{fig:maps}
\end{figure*}

\begin{figure}
\includegraphics[width=\columnwidth]{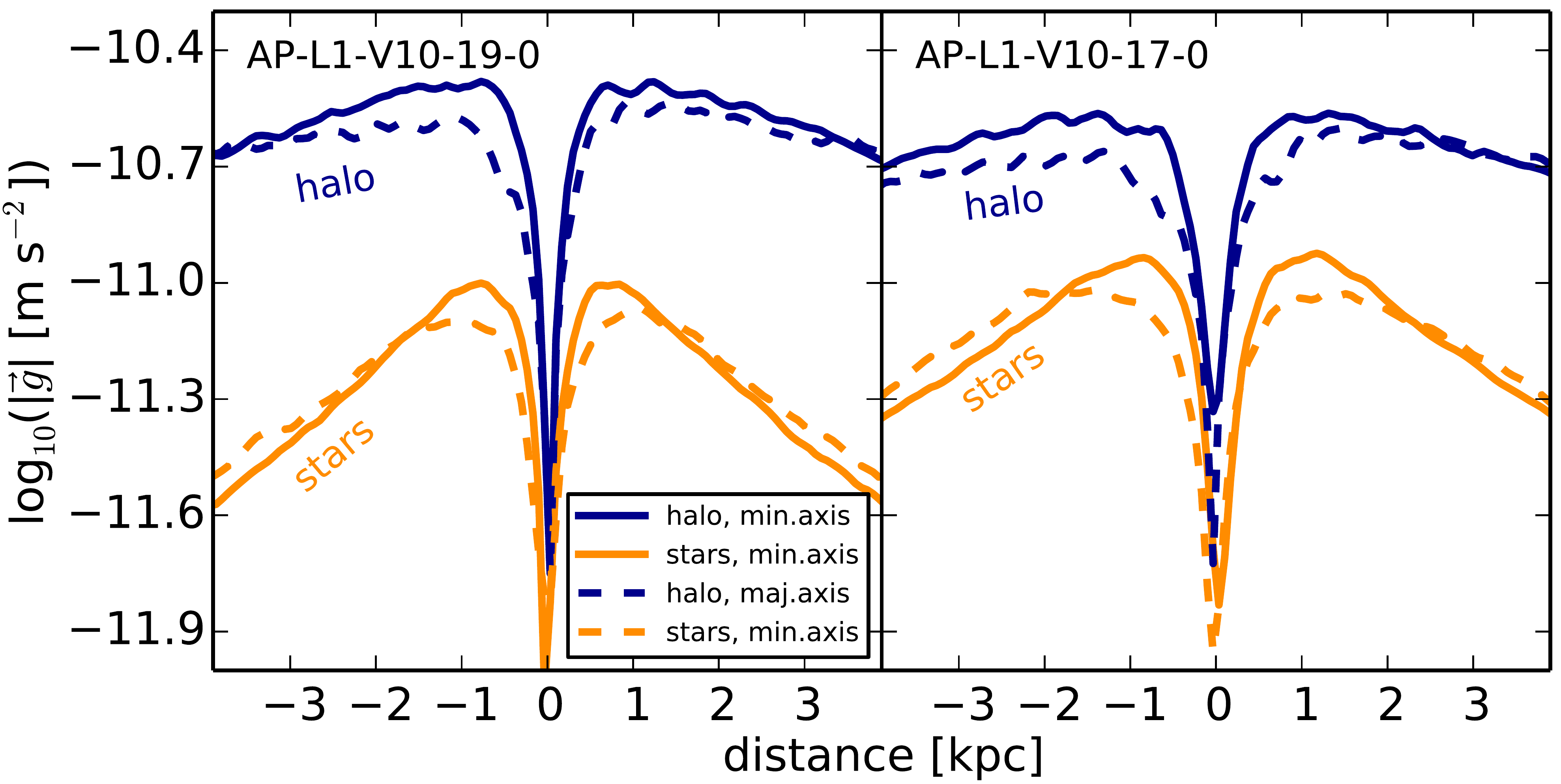}
\includegraphics[width=\columnwidth]{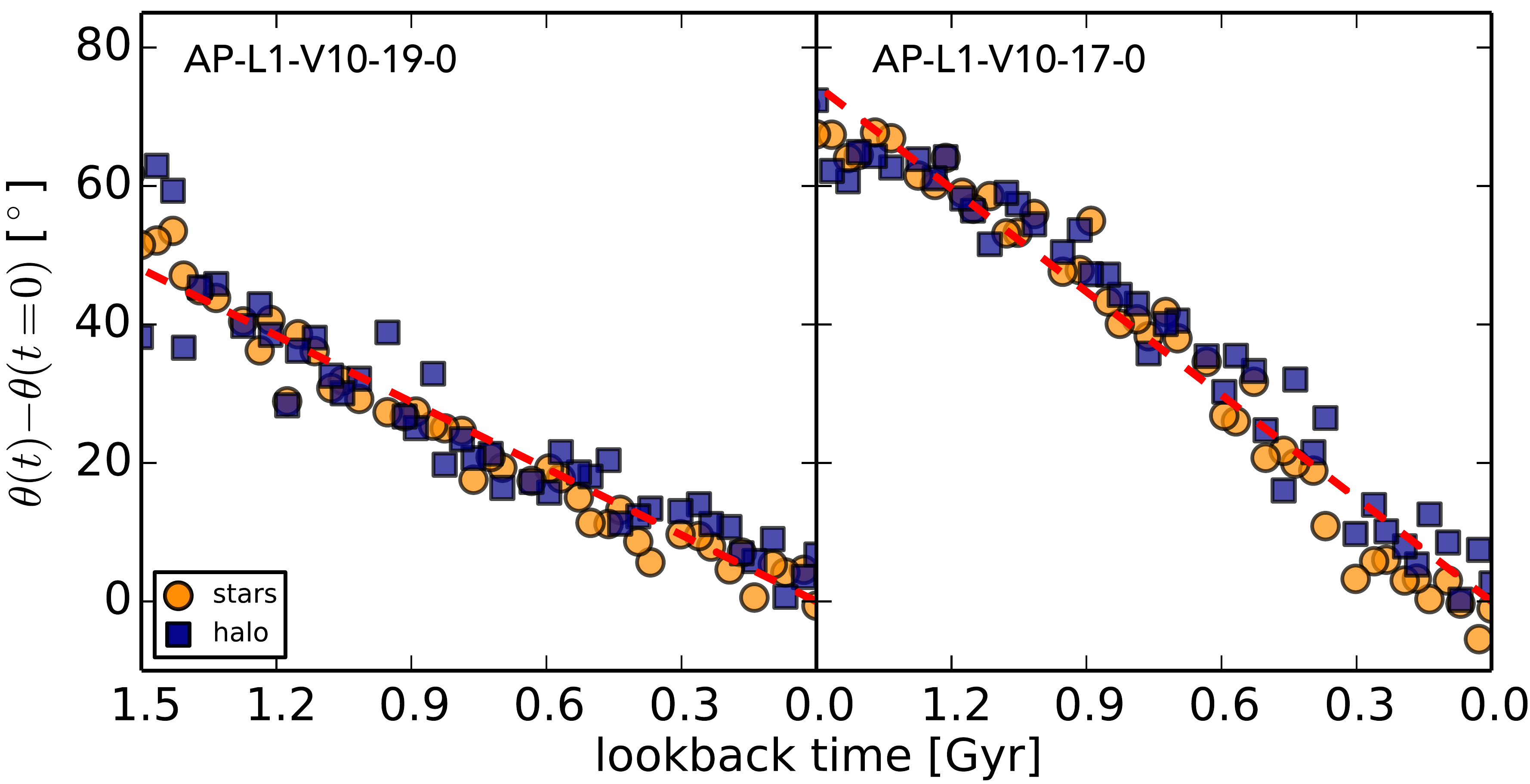}
\includegraphics[width=\columnwidth]{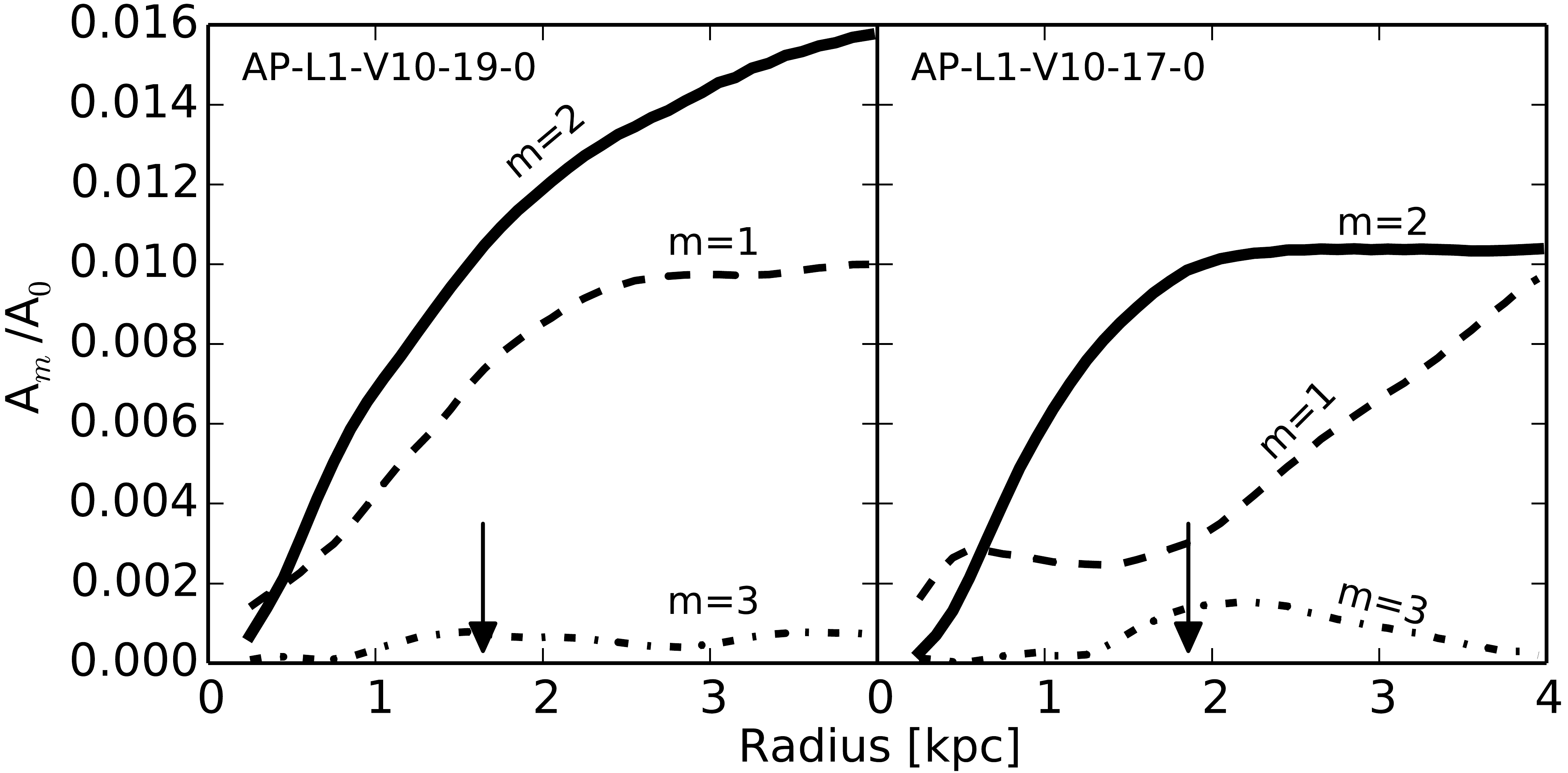}
\caption{\emph{Top panels}: gravitational acceleration on the galactic midplane decomposed into contributions from stars (orange lines) and from dark matter (blue lines) as a function of the distance along the major axis (dashed lines) or the minor axis (solid lines) of the bar for the two representative galaxies. Dark matter dominates the gravitational acceleration within the bar. \emph{Central panels}: phases of the stellar (orange circles) and of the dark matter (blue squares) bar as a function of the lookback time. Phases are locked. \emph{Bottom panels}: amplitude of the $m\!>\!0$ harmonic fluctuations in the gravitational potential $\Phi$, divided by the mean value of $\Phi$, as a function of $R$. The strongest harmonic mode is the $m\!=\!2$. The arrows show the location of $R_{\rm h}$.}
\label{fig:bar_properties}
\end{figure}

\subsection{Stellar bars and `dark bars'}

For illustration, we present first the mass distribution and the \hi\
kinematics in the inner regions of two representative systems in the
sample of O17: AP-L1-V10-19-0 and AP-L1-V10-17-0.  
These systems have comparable $V_{\rm max}$ ($67$ and $65\kms$ respectively), stellar
mass ($4.7$ and $7.1\times10^8\msun$), \hi-to-stellar mass ratio
($1.1$ and $0.7$) and star formation rates at $z=0$ ($0.08$ and
$0.13\msunyr$).  Also, both systems are quite isolated, lying at more
than $2\Mpc$ from the closest member of the Milky Way/M31 analog
pair. They also show no obvious sign of interactions with nearby
galaxies.

Fig.\,\ref{fig:maps} shows a series of face-on maps for these two
galaxies, where the rotation axis of the galaxy is identified with
$\vec{L}_*$, the angular momentum vector of all stars (identified by
{\small SUBFIND} as bound to the system and located within a radius enclosing $90$ per cent of the total stellar mass)
relative to the galaxy centre. The latter is defined as the location of the particle with the lowest potential
energy.  
All maps are smoothed to a spatial resolution (FWHM of a Gaussian kernel) of $0.2\kpc$.

The leftmost two panels of Fig.\,\ref{fig:maps} show the stellar and
dark matter surface density maps derived projecting a cubic region with
side-length $8\kpc$, concentric with the galaxy.  The
elongation of the isodensity contours reveals that \emph{both} stars
and dark matter are clearly non-axisymmetric. Projected onto the disc
plane, both components seem bar-like, especially in the central few
kiloparsecs.  For simplicity, we will refer to these non-axisymmetric
features as the `stellar' bar and the `dark' bar.

These two `bars' have the same orientation and, approximately, the same
axis ratio.  
The central panels show isopotential contours measured on the disc
midplane, computed using all (dark + baryonic) particles identified by
{\small SUBFIND} as bound to the system (the gravitational potential, $\Phi$,
is softened on scales smaller than $100\pc$).  The gravitational potential is
clearly aspherical, especially close to the centre, and is therefore
expected to induce deviations from pure circular rotation in the 
motion of gas and stars in the disc.

These deviations are explored in the rightmost two panels,
which show the azimuthal and the radial \hi\ velocity fields derived
on the galaxy midplane, from which we have subtracted the mean
azimuthally averaged value at each radius.  If the gas
was in pure circular rotation around the centre, these `residual'
velocity fields should not show systematic deviations from zero.
Instead, clear $m\!=\!2$ harmonic patterns are visible, which were
already noticed by O17 in other systems of the sample.  

The phase of the $m\!=\!2$ perturbation in the azimuthal velocity
(shown by the black crosses in the fourth column of
Fig.\,\ref{fig:maps}, which trace
the maximum positive deviation from the mean) is
oriented approximately perpendicular to the major axis of the
potential (dashed black line, see below), whereas the phase of the
perturbation in the radial velocity lags $45\de$ from the
latter.  This pattern is consistent with gas moving
counter-clockwise along elliptical orbits elongated in the same
direction of the potential.  The simplest interpretation of these
orbits is that, in these galaxies, the stellar/dark bar induces a
bisymmetric flow in the gas component. 

We determine major and minor axes for the bar by fitting an ellipse to the surface
density contour whose semi-major axis coincides with R$_{\rm h}$, the
(projected) stellar half-mass radius. This is  highlighted in blue in the leftmost panels of
Fig.\,\ref{fig:maps}.  The ellipse centre is a free parameter of the fit, in order to account for possible
offsets in the mass distribution.  $R_{\rm h}$ is computed directly from
the face-on stellar density distribution.  

\subsection{Stellar bars in sub-dominant discs}

The two leftmost panels of Fig.\,\ref{fig:maps} suggest that the dark
matter dominates the dynamics in the central few kiloparsecs. This is
shown explicitly in the top panels of Fig.\,\ref{fig:bar_properties}
which show, for both systems, the gravitational
acceleration due to the stars alone (orange lines) and to the
dark matter (blue lines) on the galactic midplane along the
projected major (dashed lines) and minor (solid lines) axes of the
bar. The gas contribution to the gravitational acceleration is negligible compared to the other components.

Fig.\,\ref{fig:bar_properties} clearly
indicates that, in these two systems, the overall dynamics is
dominated everywhere - even at the very centre - by the dark matter.
This suggests that, in these two systems, the stellar bar originates
as a response to the non-axisymmetric distribution of dark matter, and
not, as traditionally envisioned, as a result of some instability in a
stellar-dominated disc. At least some stellar bars can, therefore,
form in non-maximal discs, provided they are embedded in triaxial
dark matter halos.

One may wonder whether the alignment of the dark and the stellar
`bars' shown in Fig.\,\ref{fig:maps} is fortuitous, or if it is also
present at earlier times.  We use the \apostle\ particle data at
redshift $z\!>\!0$ to follow the evolution of these two galaxies in
the previous $1.5\Gyr$.  At each time-step, new surface density maps -
analogous to those of Fig.\,\ref{fig:maps} - are produced by
projecting the system particles for a face-on view using the same
$\vec{L}_*$ determined at $z\!=\!0$.  Note that this technique fails
if the galactic plane rotates with time, but this is not the case for
the systems considered here.

As before, we fit the $R_{\rm h}$ iso-contour of these maps with an
ellipse to determine the axis ratio and the orientation of both `bars'.
The middle panels of Fig.\,\ref{fig:bar_properties} show the phases of
the stellar (circles) and of the dark (squares) bars as a function of
the lookback time for both galaxies.  The bars are clearly locked in phase and
rotate very slowly with pattern speeds of less than $1\kmskpc$.  As we
show in Section \ref{barproperties}, this is typically the case for
barred dwarf galaxies in the \apostle\ simulations.

We can quantify the magnitude of the perturbations in the
gravitational potential induced by the bars via a harmonic analysis of
the $\Phi$ maps shown in Fig.\,\ref{fig:maps}.  For this purpose, we
divide the $\Phi$ map into a series of concentric rings of increasing
radius, each centred at the minimum of the potential and with
thickness equal to $100\pc$.  The values of $\Phi$ as a function of
the azimuthal angle $\theta$ are fit, for each ring separately, with
\begin{equation}\label{harmonic}
\Phi(\theta) = \sum_{m=0}^3 A_m \cos(m\theta - \theta_m)\,,
\end{equation} 
which is a harmonic expansion of the potential up to third order.

The bottom panels of Fig.\,\ref{fig:maps} show, as a function of $R$,
the ratio between the amplitudes $A_m$ of the $m\!>\!0$ modes and the
mean value of $\Phi$ at each radius (i.e., the $m\!=\!0$ mode
amplitude).  As expected, the strongest perturbation mode is the
$m\!=\!2$, which reaches amplitudes of $\sim0.01\times A_0$ at
$R\!=\!R_{\rm h}$ (black arrows).  Perturbations of order $m\!=\!3$
are virtually negligible, whereas $m\!=\!1$ perturbations are
intermediate.  Even though the overall fluctuations in the potential
are only of a few per cent, they are enough to affect significantly
the gas velocity field, as discussed by HN06.

We note that, while in the models of HN06 the gas follows elliptical
closed orbits with major axis oriented perpendicularly to the bar, in
the two systems studied here the major axis of the potential and that
of the gas orbits are aligned.  We find that this is the case for the
majority of our simulated dwarfs, although the alignment is not always
as clean as in AP-L1-V10-19-0 and AP-L1-V10-17-0.  

There are several differences between our galaxies and those in the
HN06 models.  The first is that the bars in our systems slowly rotate
with time, while they were steady in HN06.  The orientation of the
closed orbits in a rotating barred potential depends on the positions
of the inner Lindblad resonance and of the co-rotation radius, which
vary from case to case depending on the exact shape and pattern speed
of the potential
\citep[e.g.][]{ContopoulosGrosbol89,Athanassoula92,SellwoodWilkinson93}.
A second difference is that the HN06 models are based on the epicycle
approximation, which breaks down when perturbations in the velocity
field exceed a few per cent - they are typically of order $\sim25$ per cent at the
half-mass radius in the cases we consider here.  Finally, in our
galaxies the gas does not settle exactly on closed orbits but slowly
flows inwards with time, as we verified by following the trajectories
of a sample of gas particles in the two \apostle\ systems studied
here.  Overall, the \apostle\ dwarfs are more complex than the simple
systems considered by HN06.  A detailed analysis of the orbits of
their gas and star particles is planned for a future study.

\subsection{General properties of stellar and dark bars}
\label{barproperties}

We now extend the previous analysis to the entire sample of 33
simulated dwarf galaxies studied by O17.  These systems are selected
to have $60\!<\!V_{\rm max}/\kms\!<\!120$, and typically have
$M_{\rm HI}$ comparable to $M_*$ (see Table A1 in O17).

The dynamics of each system in this sample are dominated everywhere by
the dark matter.  This is shown in Fig.\,\ref{fig:gbar_gdark}, where
we plot the ratio of the midplane gravitational acceleration
contributed by baryons and by the dark matter, respectively, as a
function of the circular velocity, $V_{\rm circ}\def\sqrt{GM(<R)/R}$.
We show the ratios computed at $R\!=\!R_{\rm h}$ (squares) and at
$R\!=\!2R_{\rm h}$ (circles).  

A clear trend is visible: the larger $V_{\rm circ}$ is, the higher the
contribution of the baryons to the total acceleration.  However, in
all cases, the baryons are subdominant: their contribution to the
total acceleration ranges from $10\%$ to $90\%$ of that given by the
dark matter alone, decreasing towards the outer regions of the
galaxies.  Interestingly, by extrapolation of the trend shown in
Fig.\,\ref{fig:gbar_gdark}, it is clear that the baryons \emph{will}
dominate the central dynamics in galaxies more massive than those
studied here \citep[see also][]{Schaller+16}.

\begin{figure}
\centering
\includegraphics[width=\columnwidth]{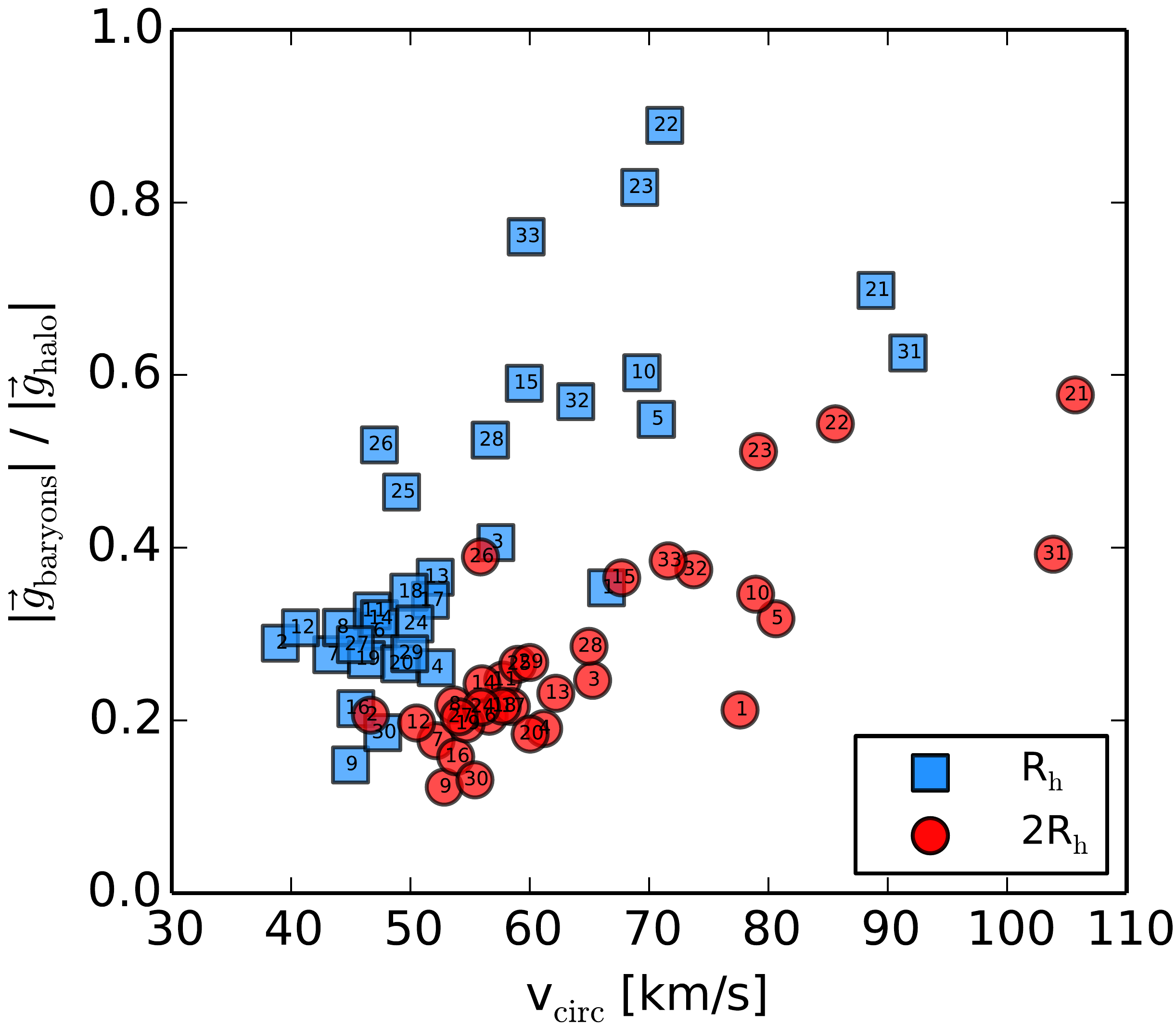}
\caption{Ratio between the baryonic and the dark matter gravitational acceleration on the midplane as a function of the circular velocity $\sqrt{GM(<R)/R}$ computed at $R\!=\!R_{\rm h}$ (squares) and at $R\!=\!2R_{\rm h}$ (circles) for the simulated dwarfs. Baryons are always subdominant.}
\label{fig:gbar_gdark}
\end{figure}

\begin{figure}
\centering
\includegraphics[width=0.9\columnwidth]{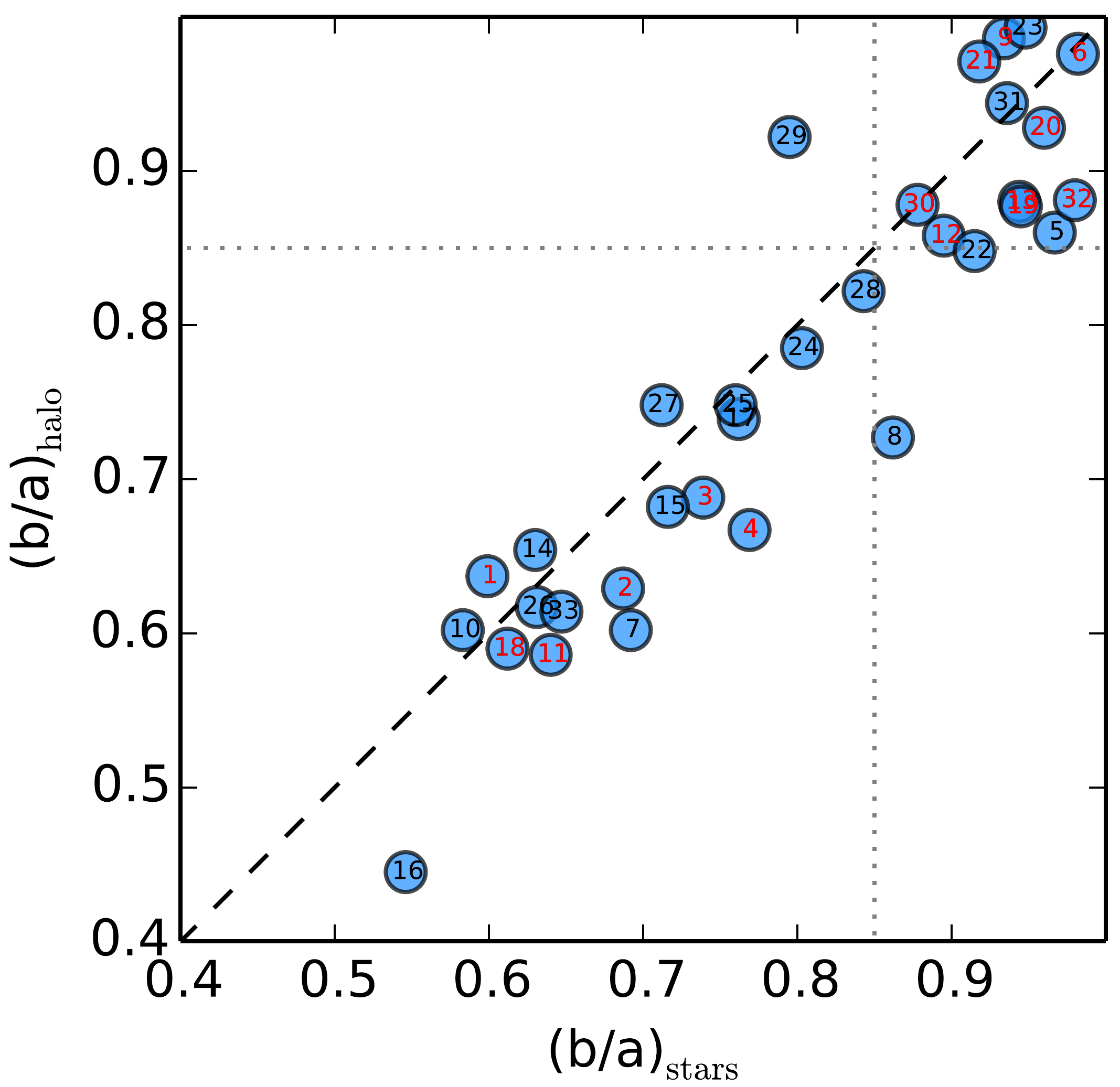}
\caption{Stellar bar versus halo bar axis ratio for our simulated dwarfs. Systems are numbered as in Table A1 of O17, with numbers in red representing galaxies `in equilibrium' (see text). $55$ per cent (18/33) of galaxies have axis ratios smaller than 0.85 and constitute our `barred' subsample.}
\label{fig:axis_ratio}
\end{figure}

Fig.\,\ref{fig:axis_ratio} compares the axis ratio of the stellar bar
with that of the dark bar for all systems in the sample.  As before,
the axis ratios are determined by fitting ellipses to the highest
contour passing by $R_{\rm h}$ in face-on surface density maps.  For
the vast majority of systems, the two axis ratios are very similar,
sugesting again that the origin of the stellar bars is closely linked
to non-axisymmetries in the dark matter. The symbol numbers in
Fig.\,\ref{fig:axis_ratio} label the various systems as in Table A1 of
O17.

We define a `barred' subsample of 18 galaxies ($55\%$ of the sample)
by selecting those systems with axis ratio smaller than 0.85 for both
the stellar and the dark matter components.  This threshold is chosen
because we found it impossible to track a bar backwards in time when
its axis ratio is rounder than 0.85.  Numbers in red in
Fig.\,\ref{fig:axis_ratio} are used to identify the subsample of 14
dwarfs identified by O17 as being `in equilibrium': these are galaxies
where the \hi\ average azimuthal speed, computed at $R\!=\!2\kpc$,
matches the circular velocity at the same radius.  In general, there
is a tendency for these systems to have a large axis ratio, suggesting
that galaxies whose halo is more spherical are, according to that
definition, closer to dynamical equilibrium than more strongly barred
systems.

\begin{figure}
\centering
\includegraphics[width=0.9\columnwidth]{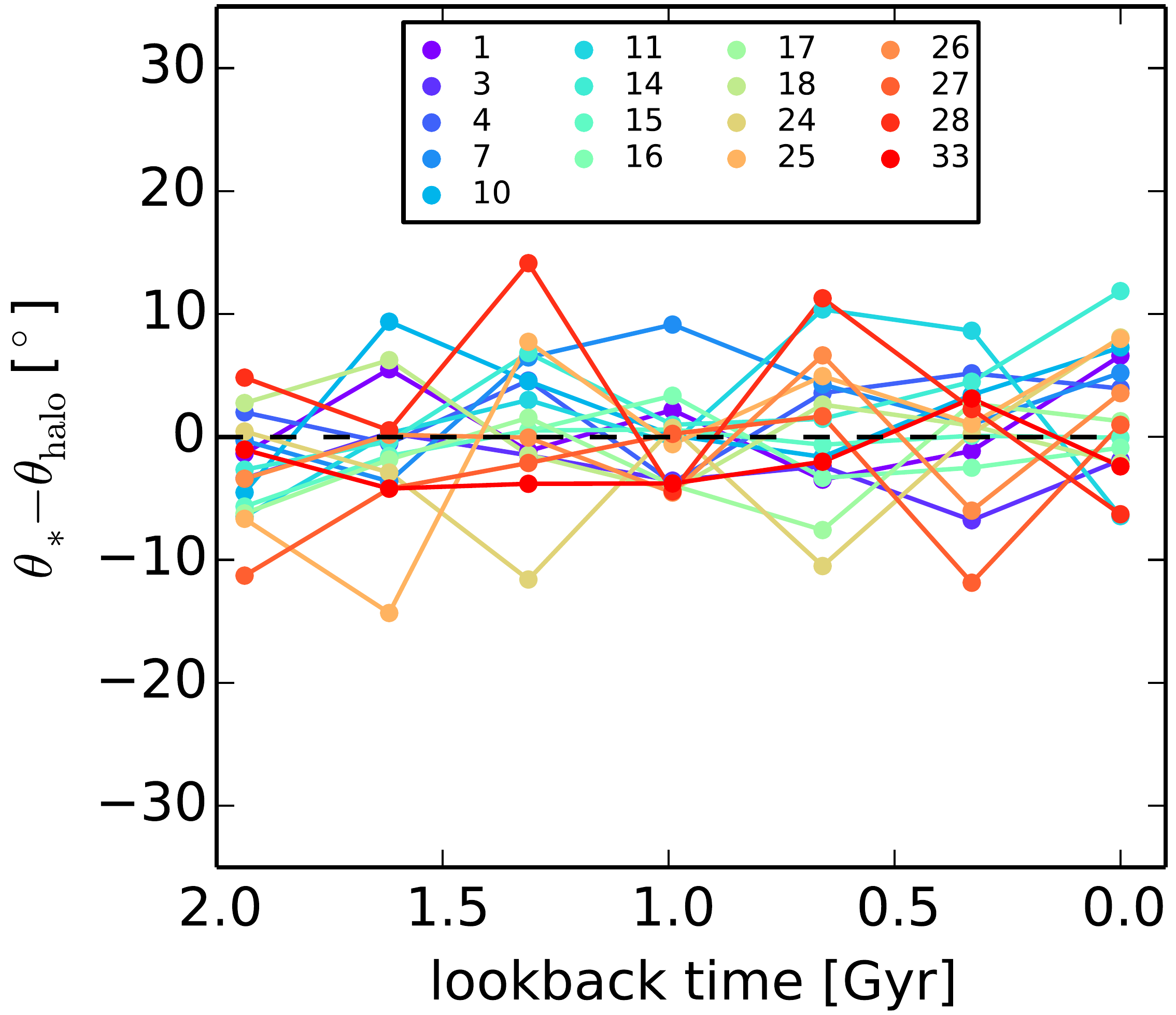}
\caption{Difference between the phase of the stellar bar and the phase of the halo bar as a function of the lookback time for our subsample of simulated barred dwarfs. Systems are numbered as in Table A1 of O17. Phases are always locked.}
\label{fig:deltaphi_vs_t}
\end{figure}

Fig.\,\ref{fig:deltaphi_vs_t} shows how the difference between the phase of the stellar bar and that of the dark bar evolves in time.
In all cases, the phase difference remains below $\sim15\de$ in the $2\Gyr$ before the present time, confirming that the stellar and dark bars always rotate in phase, and with the same pattern speed.
Note that the system labelled as `2' is missing in Fig.\,\ref{fig:deltaphi_vs_t}: a close encounter occurring at the lookback time of $\sim0.5\Gyr$ unsettled the stellar disc and made it difficult to follow the evolution of the bar at earlier epochs.

We now analyse the strength and the pattern speed of the stellar bars in our sample.
Following \citet{Algorry+17}, we define the strength of the stellar bar by measuring the amplitude of the $m\!=\!2$ Fourier mode of the azimuthal distribution of star particles. Specifically, we compute 
\begin{equation}
a_m(R) = \sum_{i=1}^{n_R} m_i \cos(m\phi_i);\!\quad b_m(R) = \sum_{i=1}^{n_R} m_i \sin(m\phi_i)
\end{equation}
where $m_i$ is the mass of the $i$-th star particle and the sum is
extended to all $n_R$ particles that occupy a given cylindrical
annulus with mean radius $R$.  We then define
$A_{2,*}(R)=\sqrt{a_2^2 + b_2^2}/a_0$, and the strength of the bar as
$A_{2,*}^{\rm max}\!=\!{\rm max}(A_{2,*}(R))$.  

\begin{figure}
\centering
\includegraphics[width=\columnwidth]{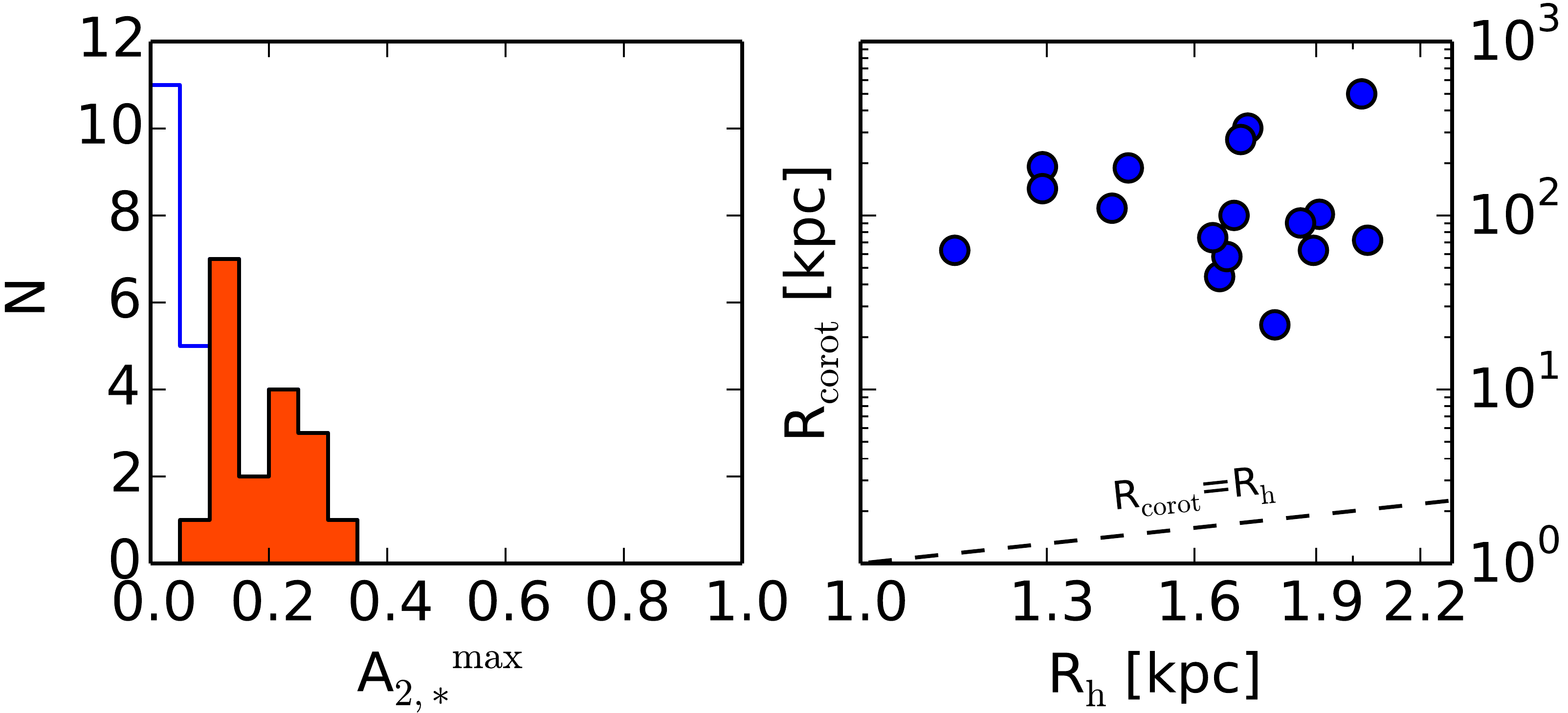}
\caption{Properties of stellar bars in our simulated dwarfs. \emph{Left panel}: strength of the bar, $A_2^{\rm max}$ (see text), for the full sample (unfilled histogram) and for the barred subsample (filled histogram). \emph{Right panel}: bar co-rotation radius plotted against the projected radius at half stellar mass. The dashed line shows the one-to-one relation.}
\label{fig:stellarbar_properties}
\end{figure}

The distribution of $A_{2,*}^{\rm max}$ is shown in the left panel of
Fig.\,\ref{fig:stellarbar_properties}.  All systems have bar strength
below $0.4$, which implies that they are all `weak bars' according to
the criterion of \citet{Algorry+17}, with only 7 systems with
$0.2\!<\!A_{2,*}^{\rm max}\!<\!0.4$.  In the right panel of
Fig.\,\ref{fig:stellarbar_properties} we compare the bar co-rotation
radius, $R_{\rm corot,*}$, and $R_{\rm h}$ in our subsample of barred
galaxies.  $R_{\rm corot,*}$ is determined by first computing the bar
pattern speed, $\Omega_*$, and then by equating $\Omega_* R$ to the
circular velocity of the system, $\sqrt{GM(<R)/R}$.  $\Omega_*$ is
computed by fitting with a straight line the trend of the phase of the
stellar bar with time, from lookback times of $2\Gyr$ to the present
day.  We find that $\Omega_*$ is typically smaller than $1\kmskpc$,
and the corresponding co-rotation radii $R_{\rm corot,*}$ always
exceed by far the system sizes.  

All stellar bars in our sample are relatively weak, very slowly
rotating, and most likely originate from the triaxial structure of
their dominant dark matter haloes.

\subsection{Baryons and halo triaxiality}
\label{barsDMO}

As discussed in Sec.~\ref{SecIntro}, the assembly of the baryonic
component of a galaxy is expected to sphericalise the dark matter
distribution. If so, this process has only gone to partial completion
in APOSTLE dwarfs, given the prevalence of non-axisymmetric features (`dark bars') in our galaxy
sample. We examine this question by studying the properties of 14
subhaloes with $60\!<\!V_{\rm max}\!<\!120\kms$ extracted from volumes
AP-L1-V1 and AP-L1-V4 in the \apostle\ dark-matter-only (DMO)
simulations, which are the DMO counterparts of the corresponding full
hydrodynamical runs.

By analogy with our previous analysis, we measure deviations from
axisymmetry by computing the axis ratio of face-on isodensity contours
with semi-major axis length of $1.7\kpc$, corresponding to the median
$R_{\rm h}$ in our original sample. Since DMO runs have no stars, we
define the face-on projection axis as the minor axis of the inertia
tensor of all dark matter particles within $8\kpc$ from the halo
centre. For consistency, we repeat the same procedure for our original
sample of 33 galaxies in the hydrodynamical runs, so that dak matter axis ratios
are derived in the same way for the two samples.

\begin{figure}
\begin{center}
\includegraphics[width=0.7\columnwidth]{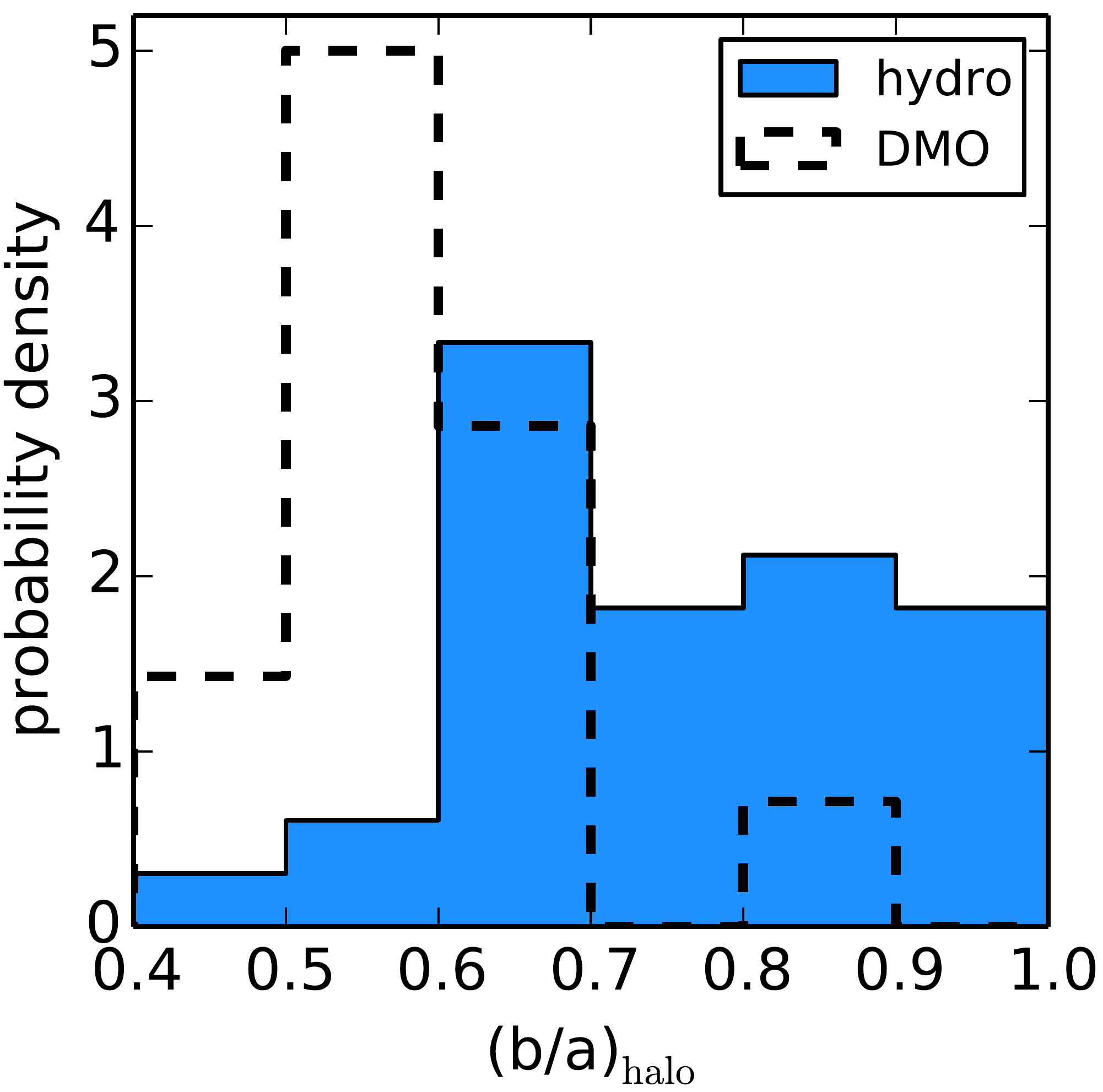}
\caption{Axis ratio distribution for the dark bars in our sample of simulated dwarfs (filled histogram) and in a sample of subhaloes with $60\!<\!V_{\rm max}\!<\!120\kms$ from the \apostle\ dark-matter-only simulations (dashed histogram). Dark bars in dark matter-only runs have lower axis ratio.}
\label{fig:axisratio_hydro_vs_dmo}
\end{center}
\end{figure}

In Fig.\,\ref{fig:axisratio_hydro_vs_dmo} we compare the $b/a$
distributions for the dark matter in the hydrodynamical (full
histogram) and in the DMO (dashed histogram) samples.  Even though the
sample is small, it is clear that deviations from axisymmetry are more
pronounced in the DMO case than in the hydro simulations.  As
expected, the assembly of baryons at the centre of a halo reduces, but
does not erase, the triaxiality of the potential in the innermost few
kpc, in line with the findings of \citet{Kazantzidis+10,Abadi+10,
  MachadoAthanassoula10}.

\subsection{Comparison with observations}
\begin{figure*}
\includegraphics[width=0.9\textwidth]{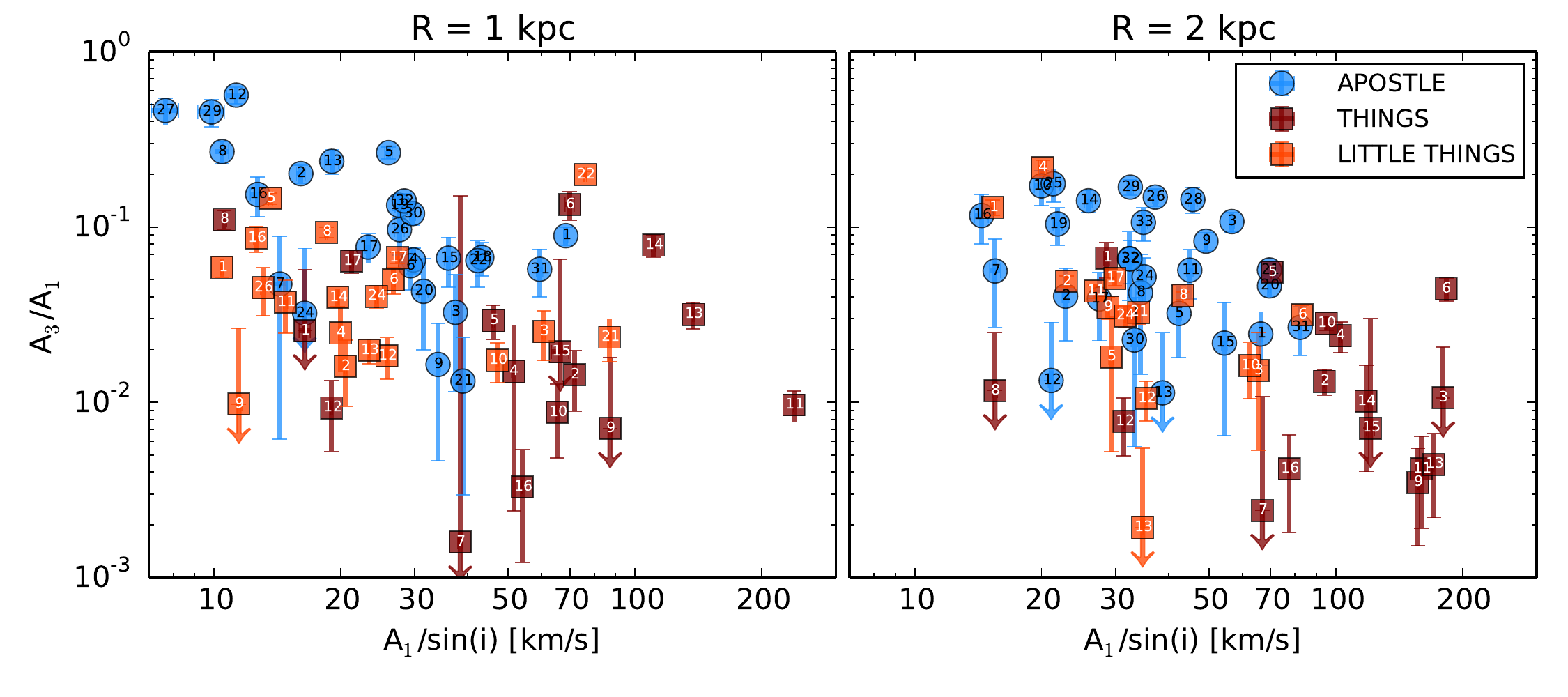}
\caption{bisymmetric \hicap\ flows in simulated and observed galaxies at galactocentric radii of $R\!=\!1\kpc$ (\emph{left panel}) and $R\!=\!2\kpc$ (\emph{right panel}). The ratio between the amplitudes of the $m\!=\!3$ and the $m\!=\!1$ harmonic modes is plotted as a function of the amplitude of the $m\!=\!1$ mode (divided by $\sin(i)$) in the \hicap\ velocity fields of \apostle\ galaxies (blue circles), THINGS galaxies (brown squares), and LITTLE THINGS galaxies (orange squares). Error bars show the formal errors on the harmonic fit to the velocities. Arrows represent upper limits. Systems are numbered as in Tables A1 and A2 of O17. On average, simulated galaxies show stronger bisymmetric motions than the observed galaxies.}
\label{fig:m3}
\end{figure*}


We now compare the amplitude of the harmonic perturbations in the \hi\
velocity fields of real galaxies with those found in our simulations.
For the latter, we use the same sample of 43 dwarfs (17 from THINGS,
26 from LITTLE THINGS) already considered by O17, adopting the
`natural'-weighted moment-1 maps of these galaxies.  These observatons
have angular resolution of $\sim12\arcsec$, corresponding to a median
spatial resolution of order $\sim250\pc$.  The overall kinematics of
these systems has been already studied and reported by \citet{deBlok+08}
and \citet{Oh+15}.

In observed galaxies we do not have access to the 3D velocity field,
but only to its projection along the line-of-sight.  As is well known
\citep[e.g.][]{Schoenmakers+97, SpekkensSellwood07}, harmonic
perturbations of order $m$ in the azimuthal or radial velocity field
produce harmonic distortions of orders $m'\!=\!m\pm1$ in the
line-of-sight velocity field.  Hence a bisymmetric flow in real space
produces an $m'\!=\!1$ \emph{and} an $m'\!=\!3$ perturbations in
projected space, both having similar amplitude.

While the former is difficult to detect, as its signal blends with that produced by the regular rotation of the disc, 
the latter is a unique signature of bisymmetric flows
and can be readily studied via a harmonic decomposition of the
velocity field.  Our strategy is therefore to produce synthetic
line-of-sight \hi\ velocity fields for the simulated systems, with
resolution similar to that of observed galaxies, and compare the
amplitudes of the harmonic modes of simulated and observed galaxies.

In order to produce synthetic \hi\ datacubes for the 33 systems in our simulated sample, we follow a procedure similar to that described by O17 in their Section 3.3.
The procedure consists of the following steps.
Assume that a system is made of $N$ gas particles, each described by its coordinates $(x,y,z)$, velocities $(V_x,V_y,V_z)$, temperature $T$, and \hi\ mass $m_{\rm HI}$:
\begin{itemize}
\item the system is projected at $60\de$ inclination, using
  $\vec{L}_*$ as a reference direction, and a random orientation;
\item a position-velocity 3D grid is created, with a spatial binning of $83\pc$ and a channel separation of $2\kms$;
\item the $m_{\rm HI}$ of each particle is placed at the corresponding $(x,y)$ location in the grid, and is distributed in velocity around its $V_z$ by using a Gaussian kernel with a standard deviation of $\sqrt{k_{\rm B}T/m_{\rm H}}$, $k_{\rm B}$ being the Boltzmann constant and $m_{\rm H}$ the proton mass. This is to take the thermal broadening of the line profiles into account.
\item once all particles have been processed, the resulting datacube is smoothed spatially at the FHWM resolution of $250\pc$ (three times the grid binning).
\end{itemize}

For simplicity we make no correction for gas opacity, i.e., the gas is considered to be optically thin.
Note that, as in O17, the final resolution of our synthetic
observations is similar to that of THINGS and LITTLE THINGS datacubes.
Velocity fields are derived from these datacubes as moment-1
maps, which are well suited to account for non-circular motions in the
gas kinematics. 

Finally, we perform a harmonic analysis of the velocity fields for both samples.
The procedure requires a careful choice of galaxy centre.  For the
simulated galaxies, we assume that this coincides with the location of
the minimum gravitational potential, whereas for the real galaxies we
take the kinematic centres estimated by \citet{Trachternach+08} for
THINGS and by \citet{Oh+15} for LITTLE THINGS.  

We focus on the amplitude of the $m\!=\!1$ and $m\!=\!3$ harmonic
modes at two fixed galactocentric distances: $R\!=\!1\kpc$ and
$R\!=\!2\kpc$, which bracket the range of $R_{\rm h}$ in our simulated
sample and are well-resolved radii in both observations and
simulations. In practice, we select a ring
on the galactic plane with mean radius equal to the chosen value ($1$
or $2\kpc$) and width equal to the FWHM resolution (which varies in
the sample of real galaxies, but is constant in the simulated sample).
Each ring is defined by its inclination, $i$, and position angle,
$\rm PA$, in the sky.  

Given that the choice of these two parameters
significantly affects the outcome of the harmonic analysis, we must
adopt a criterion that sets them uniquely and that can be applied to
both observed and synthetic velocity fields.  For any given ($i$,
$\rm PA$), the line-of-sight velocity, $\vlos$, within the ring as a
function of the azimuthal angle in the plane of the galaxy, $\theta$,
is fit with a formula analogous to eq.\,(\ref{harmonic}), from which
we extract the amplitudes $A_1$, $A_2$ and $A_3$.  Our choice for
($i$, $\rm PA$) is the one that \emph{minimises} the quantity
$\sqrt{A_2^2+A_3^2}/|A_1|$, which quantifies the strength of
large-scale perturbations with respect to regular rotation at a given
radius. The minimisation is achieved via the \citet{NelderMead65}
method.  

We adopt ($i\!=\!60\de$, ${\rm PA}\!=\!90\de$) as the initial
guess for the simulated galaxies, while for the observed galaxies we
use the values of $i$ and $\rm PA$ determined by \citet[][for
THINGS]{deBlok+08} and by \citet[][for LITTLE THINGS]{Oh+15} from
their tilted ring fitting method.  Typically, the final ($i$,
$\rm PA$) determined for the observed galaxies remains within
$\sim 5\de$ from the initial estimates.  The simulated sample,
however, shows much larger deviations from the initial guess.

In some cases, fitting $\vlos$ via eq.\,(\ref{harmonic}) was not
possible due to the small number of points, or - in observed galaxies
- to an excessively noisy velocity field.  For these reasons DDO 53,
DDO 210, IC 10, IC 1613, NGC 1569, UGC 8508 and Haro 29 have been
excluded from our analysis at $R\!=\!1\kpc$, along with DDO 168, DDO
216, NGC 3738 and Haro 36 at $R\!=\!2\kpc$.


Figure \ref{fig:m3} shows the ratio between $A_3$ and $A_1$, which is
a measure of the strength of the bisymmetric flow with respect to
regular \hi\ rotation, as a function of $A_1/\sin(i)$, which
represents the \hi\ azimuthal speed $V_\phi$ alone\footnote{In
  reality, $A_1$ is a combination of rotation and a global
  expansion/contraction in the radial direction, but the latter is
  typically negligible}, at $R\!=\!1\kpc$ (left panel) and
$R\!=\!2\kpc$ (right panel) for the real (squares) and simulated
(circles) galaxies.

Interestingly, there is a common trend followed by both real and
simulated galaxies: the importance of non-circular motions compared to
the regular rotation decreases as a function of rotation speed.  The
main difference between observed and simulated galaxies is in the
amplitudes of the bisymmetric motions: the median $A_3/A_1$ at
$R\!=\!1\kpc$ ($R\!=\!2\kpc$) is $0.083$ ($0.057$) in \apostle,
$0.017$ ($0.010$) in THINGS and $0.040$ ($0.033$) in LITTLE THINGS.
Limiting the comparison to the LITTLE THINGS sample alone, which spans
a range of $A_1/sin(i)$ similar to our \apostle\ sample, we conclude
that bisymmetric flows in \apostle\ dwarfs are a factor $\sim2$
stronger than those in the observed galaxies.

We have verified that these results hold when the harmonic analysis is
performed at the effective radius $R_{\rm eff}$ of each system, or at
$2R_{\rm eff}$, rather than at a fixed galactocentric distance, using
effective radii from the SPARC catalogue of \citet{Lelli+16} and the
catalogue of \citet{HunterElmegreen06}.

These results suggest that bisymmetric flows caused by bar-like
features, or, more generally, by asphericities in the gravitational
potential are somewhat less prominent in observed galaxies than in the
\apostle\ simulations, at least in the dwarf galaxy regime, in line
with the earlier findings of \citet{Trachternach+08}.


There are, however, important caveats in the comparison. One is that,
while our analysis considers all \apostle\ galaxies in the range
$60\!<\!V_{\rm max}/\kms\!<\!120$, the observed sample has no
well-defined completeness criteria. A second difference, already
mentioned above, is in the velocity range covered: only 8 galaxies in
the observed sample have $60\!<\!V_{\rm max}/\kms\!<\!120$.  A further
difference is environmental: the \apostle\ galaxies are located in the
proximity (i.e. within $\sim3\Mpc$) of a Local Group analogue, whereas
the THINGS and LITTLE THINGS galaxies are in less dense environments.
In light of these considerations, we argue that the differences in the
$A_3/A_1$ ratio shown in Figure \ref{fig:m3}, although suggestive,
should be treated with caution.

\section{Conclusions}
\label{Conclusions}

Cold dark matter haloes are triaxial in nature
\citep[e.g.][]{Frenk+88}.  Their asphericity increases towards their
centre \citep{Hayashi+07}, implying that the process of galaxy
formation in a $\Lambda$CDM framework occurs within gravitational
potentials that are non axisymmetric, or, broadly speaking, `barred'.
The processes of stellar mass assembly and secular evolution can
reduce the halo triaxiality significantly in massive disc galaxies
\citep{Kazantzidis+10,Abadi+10,MachadoAthanassoula10}, but are much
less efficient in dwarf galaxies, which therefore may be forced to
respond to non-axisymmetric forces due to the dark matter.

In this work, we have carried out a study of the mass distribution and \hi\ kinematics within the central few kiloparsecs in a sample of 33 \hi-rich dwarf ($60\!<\!V_{\rm max}\!<\!120\kms$) galaxies from the \apostle\ suite of $\Lambda$CDM cosmological hydrodynamical simulations.
Our results can be summarised as follows:
\begin{itemize}
\item Most simulated dwarfs have a bar in their stellar component that
  matches the non-axisymmetric distribution of the inner dark matter
  halo (`dark' bar).  Specifically, 18 out of 33 \apostle\ dwarfs ($55$ per
  cent) have a stellar and a dark bar with axis ratios smaller than
  0.85.
\item When present, the stellar and the dark bar co-rotate and are locked in phase.
\item Since the gravitational potential is dominated by the dark matter, the stellar bar follows the dark bar, not viceversa. 
A corollary is that the presence of a stellar bar does not imply that baryons dominate gravitationally that region.
\item All bars in the sample analysed are weak, and have co-rotation radii that largely exceed the galaxy size.
\item The stellar/dark bar induces significant bisymmetric flows in
  the gas component. We have compared the amplitude of the $m\!=\!3$ harmonic perturbations in the \hi\ velocity fields of the \apostle\ systems with those of THINGS and LITTLE THINGS galaxies, finding that the magnitude of bisymmetric flows in the former exceeds that in the latter by a factor of $\sim2$.
\end{itemize}

Our findings clarify the nature of the non-circular motions reported
by O17 in the same simulated galaxy sample as due to the effects of a
dominant triaxial halo. Our analysis also shows that, although halos
are sphericalised by the assembly of the galaxy, the sphericalisation
is incomplete in dwarfs like the ones in our \apostle\ sample. The
remaining triaxiality induces the formation of a bar-like feature in
the stars and non-circular motions in the gas. Although the magnitude
of such motions seems to exceed, on average, those in galaxies of the
THINGS and LITTLE THINGS surveys, we caution that these surveys are
not exactly comparable, so a definitive conclusion about whether our
findings are in agreement or disagreement with real galaxies in the
local Universe remains pending and will be the focus of future work.

\section*{Acknowledgements}
The authors thank the THINGS, LITTLE THINGS and SPARC survey teams for
making their data publicly available.  AM thanks Mattia Sormani for
helpful discussions.  CSF acknowledges support from ERC Advanced Grant
267291 Cosmiway.  This work used the DiRAC Data Centric system at
Durham University, operated by the Institute for Computational
Cosmology on behalf of the STFC DiRAC HPC Facility (www.dirac.ac.uk).
This equipment was funded by BIS National E-infrastructure capital
grant ST/K00042X/1, STFC capital grant ST/H008519/1, and STFC DiRAC
Operations grant ST/K003267/1 and Durham University.  DiRAC is part of
the National E-Infrastructure.




\bibliographystyle{mnras}
\bibliography{marasco_bars} 





\bsp	
\label{lastpage}
\end{document}